\documentclass[twocolumn,fontsize=9pt]{scrartcl}
\usepackage[a4paper, top=0.5in,left=0.5in,right=0.5in,bottom=0.8in]{geometry}
\usepackage{graphicx}
\usepackage{lmodern}
\usepackage[utf8]{inputenc}
\usepackage{amsmath}
\usepackage[usenames,dvipsnames,svgnames,table]{xcolor}
\usepackage{float}
\usepackage[tableposition=top,labelfont=bf,textfont=sf,format=plain]{caption}
\usepackage{subcaption}
\usepackage{tabularx}
\usepackage{nicefrac}
\usepackage{siunitx} 
\usepackage{hhline}
\usepackage{filecontents}
\usepackage{layouts}

\definecolor{darkblue}{rgb}{0,0,.3}
\usepackage[bookmarksopen=true, bookmarksopenlevel=4, breaklinks=true, pdfstartview=Fit,colorlinks=true,
			linkcolor=darkblue, 
			citecolor=darkblue, 
			urlcolor=darkblue, 
			filecolor=darkblue]{hyperref}

\newcommand{\amu}{\,\si{\atomicmassunit}}
\newcommand{\pamu}{\,\si{\pico\atomicmassunit}}
\newcommand{\mum}{\,\si{\micro\metre}}
\newcommand{\kHz}{\,\si{\kHz}}
\newcommand{\MHz}{\,\si{\MHz}}
\newcommand{\mHz}{\,\si{\mHz}}
\newcommand{\Hz}{\,\ensuremath{\mathrm{Hz}}}
\newcommand{\K}{\,\si{\K}}
\newcommand{\tesla}{\,\si{\tesla}}
\newcommand{\kelvin}{\,\si{\kelvin}}
\newcommand{\mbar}{\,\si{\milli\bar}}
\newcommand{\mubar}{\,\si{\micro\bar}}

\newcommand{\HDp}{\isotope{HD}^+}
\newcommand{\Csp}{\isotope[12]{C}^{6+}}
\newcommand{\Cfp}{\isotope[12]{C}^{4+}}
\newcommand{\secs}{\,\si{\second}}
\newcommand{\eV}{\,\si{\eV}}
\newcommand{\meV}{\,\si{\milli \eV}}
\newcommand{\meter}{\,\si{\meter}}

\title{Penning-trap mass measurements of the deuteron and the $\HDp$ molecular ion}
   \small \author{\parbox{15cm}{Sascha Rau$^{1,*}$, Fabian Heiße$^{1,2}$, Florian Köhler-Langes$^1$, Sangeetha Sasidharan$^{1,2}$, Raphael Haas$^{2,3,4,5}$, Dennis Renisch$^{3,4}$, Christoph E. Düllmann$^{2,3,4,5}$,
   Wolfgang Quint$^2$, Sven Sturm$^1$, Klaus Blaum$^1$}}
\date{%
    $^1$Max-Planck-Institut für Kernphysik, 69117 Heidelberg, Germany\\
$^2$GSI Helmholtzzentrum für Schwerionenforschung GmbH, 64291 Darmstadt, Germany\\
$^3$Johannes Gutenberg-Universität, 55099 Mainz, Germany\\
$^4$Helmholtz-Institut Mainz, 55099 Mainz, Germany\\
$^5$PRISMA+ Cluster of Excellence, Johannes Gutenberg-Universität Mainz, 55128 Mainz, Germany\\
$^*$e-mail: sascha.rau@mpi-hd.mpg.de
}
\usepackage{isotope}

\begin{document}
\definecolor{orangemovement}{RGB}{225,128,0} 
\maketitle

\textbf{
The masses of the lightest atomic nuclei and the electron mass \cite{Sturm2014} are interlinked and are crucial in a wide range of research fields, with their values affecting observables in atomic \cite{Myers2019}, molecular \cite{Biesheuvel2016,Korobov2017,Alighanbari2020} and neutrino physics \cite{Otten2008} as well as metrology.
The most precise values for these fundamental parameters come from Penning-trap mass spectrometry, which achieves relative mass uncertainties in the range of $10^{-11}$. However, redundancy checks using data from different experiments reveal significant inconsistencies in the masses of the proton ($m_p$), the deuteron ($m_d$) and helion ($m_\text{he}$), amounting to $5$ standard deviations for the term $\Delta=m_p+m_d-m_{\text{he}}$,
which suggests that the uncertainty of these values may have been underestimated.
Here we present results from absolute mass measurements of the deuteron and the $\HDp$ molecular ion against $\isotope[12]{C}$ as a mass reference.
Our value for the deuteron
$m_d=2.013\,553\,212\,535 (17)\amu$
supersedes the precision of the literature value \cite{CODATA2018} by a factor of $2.4$ and deviates from this by $4.8$ standard deviations. With a relative uncertainty of $8$ parts per trillion (ppt) this is the most precise mass value measured directly in atomic mass units.
Furthermore, the measurement of the $\HDp$ molecular ion,
$m(\HDp)=3.021\,378\,241\,561\,(61)\amu$,
not only allows for a rigorous consistency check of our measurements of the masses of the deuteron (this work) and proton \cite{Heisse2019}, but also establishes an additional link for the masses of tritium \cite{Myers2015} and $\text{helium-3}$ \cite{Hamzeloui2017} to the atomic mass unit.
Combined with a recent measurement of the deuteron-to-proton mass ratio \cite{Myers2020} the uncertainty of the reference value of $m_p$ [8] can be reduced by a factor of three.
This is a post-peer-review, pre-copyedit version of an article published in Nature. The final authenticated version is available online at \url{https://doi.org/10.1038/s41586-020-2628-7}
}

\begin{figure}[tb]
\includegraphics[width=0.5\textwidth]{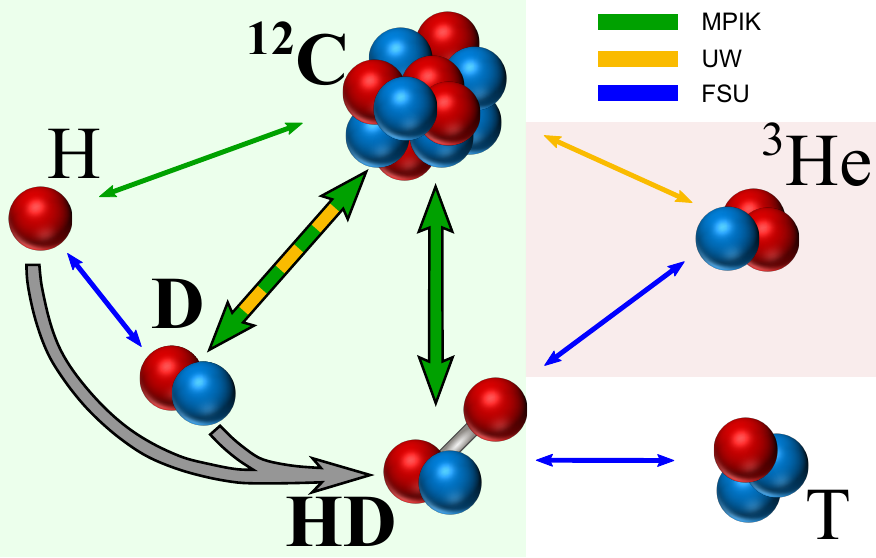}
\caption{\textbf{Overview of cyclotron frequency ratio ($R^{\text{CF}}$) measurements on light ions.}
The double-headed arrows indicate measurements color coded by group with the values reported in this work drawn thicker.
The mass of the $\isotope{HD}$ molecule can be derived using the individual masses of $\isotope{H}$ and $\isotope{D}$ and the molecular binding energy, which is sufficiently well known (grey arrow). Using the measurements of our group (MPIK), the measurements with green shaded background are self-consistent, which is no longer the case when additionally considering the measurements relating to $\isotope[3]{He}$ in the red shaded area. The link between $\isotope[3]{He}$  and $\isotope{T}$ is currently derived from individual measurements linking these ions to $\isotope{HD}$ and is of interest for the KATRIN experiment. For details and references see text.
}
\label{fig:puzzle}
\end{figure}

Penning traps allow the extremely precise determination of the ratio of atomic masses.
This way, over the years an extensive network has been created that connects individual masses via one or several links (see \autoref{fig:puzzle}).
The recently implemented redefinition of the international system of units (SI) \cite{SI2019} generally allows expressing atomic masses in $\si{\kilo\gram}$ with relative uncertainties of $3\times 10^{-10}$ \cite{CODATA2018}.
However, especially for fundamental physics applications a direct and trustworthy link from the lightest ions to the atomic mass unit $\si{\atomicmassunit}$ (one twelfth of the mass of a \isotope[12]{C} atom) with higher precision is desirable to enable e.g. a connection to the electron mass, which is measured in $\si{\atomicmassunit}$. 
However, different links between the masses of the proton ($m_p$), the deuteron ($m_d$) and the helion (the $\isotope[3]{He}$ nucleus $m_{\text{he}}$) show a discrepancy of about $5$ standard deviations, questioning the reliability of the tabulated values for these important masses. To see this, we examine $\Delta=m_p+m_d-m_{\text{he}}$. This value, which is related to the proton separation energy of  $\text{helium-3}$, can be derived in two ways
either by using the measurements relating the involved masses directly to $\isotope[12]{C}$ \cite{Heisse2019,Zafonte2015} ($\Delta_{\isotope{C}}$), or by a mass ratio measurement of the $\HDp$ molecular ion and the $\isotope[3]{He}^+$ ion \cite{Hamzeloui2017} ($\Delta_{\text{FSU}}$).
Comparing both values using the measurements available prior to this work yields a difference $\Delta_{\isotope{C}}-\Delta_{\text{FSU}}=484\,(97)\pamu$.
Here, $\pamu$ denotes $10^{-12}\amu$ and the number in brackets denotes one standard deviation. In the atomic mass evaluation 2016 (AME16) \cite{AME2016}, this problem was treated by omitting the direct $\text{helium-3}$ measurement from the adjustment process, while still using the deuteron mass reported by the same group in the same measurement campaign, potentially leading to underestimated errors in the adjustment. In \autoref{fig:puzzle} we give an overview over the light ion mass measurements, also including the measurements reported in this work and a recent measurement of the deuteron-to-proton mass ratio \cite{Myers2020}. Note that the conversion between the mass of ions and the mass of the corresponding atoms is possible without loss in precision in the low mass regime.

In Penning-trap mass ratio measurements, the cyclotron frequency $\nu_c=\frac{1}{2\pi} \frac{q}{m}B$ of a single ion with charge $q$ and mass $m$ in a homogeneous magnetic field $B$ is compared to the cyclotron frequency of a reference ion in the same magnetic field. Using $\Csp$ as reference, the mass of deuteron is then given by
\begin{equation}
m_d=\frac{1}{6}\frac{\nu_c \left( \isotope[12]{C}^{6+}\right )}{\nu_c \left (d\right) } m\left ( \isotope[12]{C}^{6+} \right)=\frac{1}{6}R^{\text{CF}}m\left(\isotope[12]{C}^{6+}\right).
\label{eq:eq1}
\end{equation}
The atomic mass of the highly charged carbon ion is known very well, since the electron's atomic mass and the ionization energies \cite{NIST_IP} are known with sufficient precision: $m(\Csp)=11.996\,709\,626\,412\,46\,(35)\amu$ with a relative uncertainty of $0.03$ ppt. The experiment therefore comes down to measuring the cyclotron frequency ratio $R^{\text{CF}}$ as precisely as possible.

\textbf{Setup}

The measurements described here are carried out in the LIONTRAP apparatus, a cryogenic Penning-trap mass spectrometer dedicated to light ion mass measurements \cite{Heisse2019}. The setup consists of a stack of Penning traps, including a highly optimized seven-electrode precision trap (PT) and two adjacent storage traps (ST), located in the homogeneous field of a superconducting $B_0=3.8\tesla$ magnet. Further details of our setup are described elsewhere \cite{Heisse2019,Heisse2017}.
In a Penning trap, a superposition of a homogeneous magnetic field in axial direction and an electrostatic quadrupolar potential confines the ion's motion. In the axial direction, the ion performs harmonic oscillations with frequency $\nu_z\approx460\kHz$. The motion in radial direction splits into two independent eigenmotions, the modified cyclotron motion with frequency $\nu_+\approx30\MHz$ and the magnetron motion with frequency $\nu_-\approx4\kHz$. These values are approximate numbers for our setup and particles with charge-to-mass ratio $\frac{q}{m}\approx\frac{1}{2}\frac{\si{\elementarycharge}}{\si{\atomicmassunit}}$.
The free cyclotron frequency is related to the motional frequencies by an invariance theorem $\nu_c=\sqrt{\nu_+^2+\nu_z^2+\nu_-^2}$, which is invariant with respect to tilts and elliptical deformations of the electrostatic potential \cite{Brown1982}. The detection principle is based on image currents induced in the electrodes by the axial oscillation of the ion. These currents are in the order of $10^{-15}\si{\ampere}$ and get transimpedance amplified into measurable voltages by a superconducting tank circuit in resonance with the axial motion and a cryogenic low-noise amplifier. The interaction with the tank circuit provides a heat sink, which brings the ion’s axial temperature into equilibrium with the ambient temperature ($\approx 4.2\kelvin$), or even lower by means of electronic feedback (FB) cooling \cite{Urso2003}.
The axial frequency is determined by a fit to the lineshape of the amplified and Fourier transformed thermal noise of the resonator, where the thermalized ion appears as a short in the noise spectrum called a “dip”.
In the invariance theorem used to extract the free cyclotron frequency, the modified cyclotron frequency has the highest significance because of the strong hierarchy between the motional frequencies
 $\nu_+\gg \nu_z\gg \nu_-$. Therefore, we employ the phase-sensitive Ramsey-like measurement method PnA (``Pulse `n' Amplify'')\cite{Sturm2011} to measure $\nu_+$ with highest precision. In the PnA method, a dipolar pulse excites the modified cyclotron mode, imprinting a phase. After some phase-evolution time, the phase is read out in the axial motion by applying a quadrupolar pulse on the sideband $\nu_++\nu_z$,
which amplifies the modified cyclotron as well as the axial motion parametrically and transfers the phase of the modified cyclotron motion to the axial motion.
This method is lineshape independent, but one has to correct for relativistic frequency shifts due to the excited radius during the phase-evolution.
The magnetron frequency has been measured at various times throughout the data taking using the ``double-dip'' technique as in our previous campaign \cite{Heisse2019}. For this, the drive-strength is varied and extrapolated to zero.

\begin{figure*}[tb]
	\centering
	\includegraphics[width=\textwidth]{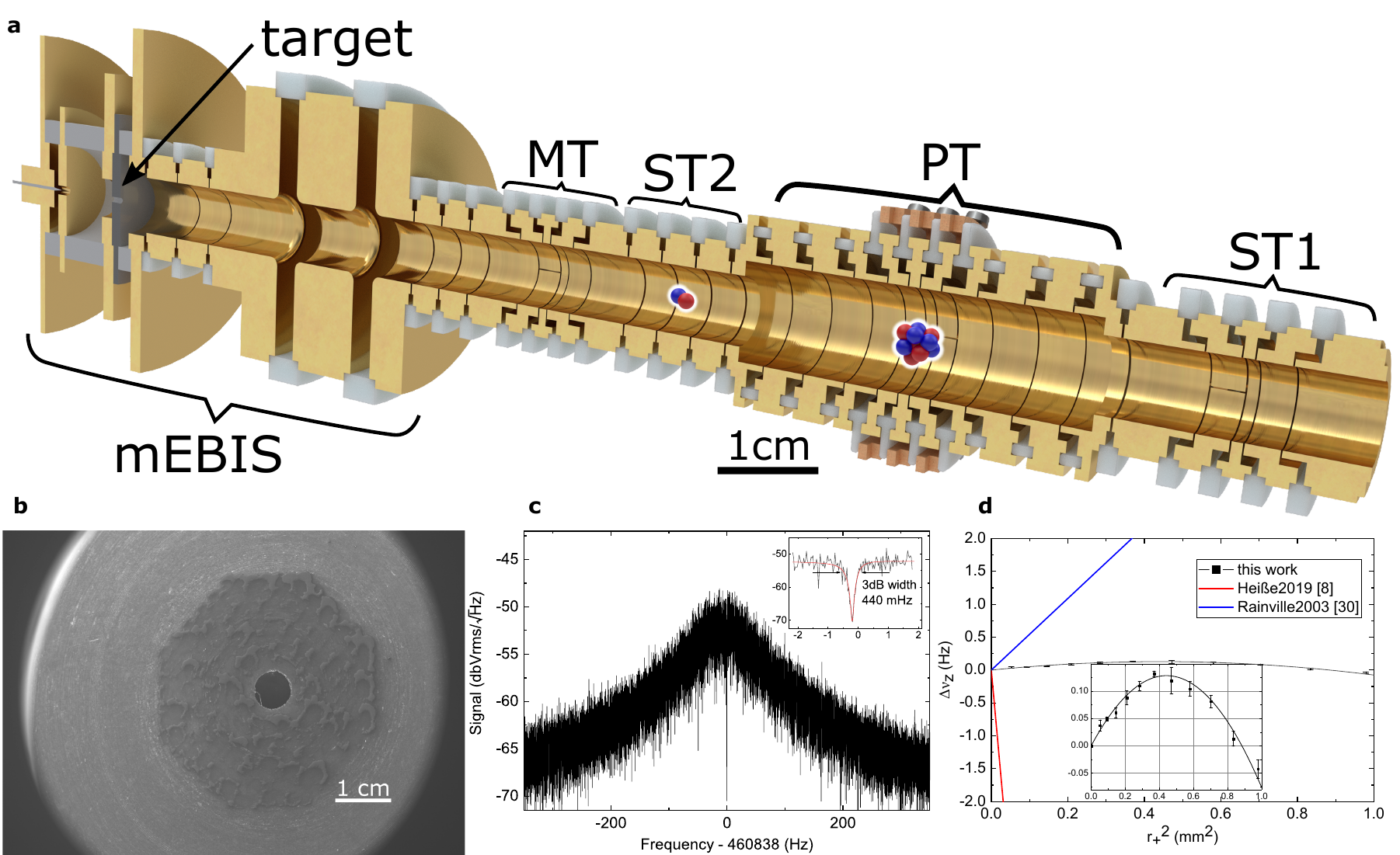}
	\caption{\textbf{Details of the LIONTRAP setup.}
\textbf{(a)} Sectional sketch of the Penning-trap stack. The ions of interest are loaded at the same time within the trap setup and measured
alternately in our precision trap (PT), while the other ion is stored in a neighboring storage trap (ST). The magnetometer trap (MT) was not used in the present work.
A miniature electron beam ion source (mEBIS) is used for the production of ions.
The trap tower is oriented vertically in our setup, with the ST1 at the top.
\textbf{(b)} Electron microscope image of the target. The printed layer of deuterated molecules used for producing deuterium ions is visible as octagonal darker area. Accelerated electrons pass through a $700\mum$ hole in the center, are reflected back and hit the deuterated area.
\textbf{(c)} Axial "dip" signal of a single deuteron. The inset shows a zoom of the central region together with a fit function.
\textbf{(d)} Measurement of the residual magnetic bottle component $\nicefrac{B_2}{B_0}$ of our magnetic field after shimming. The amplitude of the frequency shift corresponding to a $\nicefrac{B_2}{B_0}$ as in our previous experiment \cite{Heisse2019} (red) and as in an experiment, where in-situ shimming using the magnet's shim coils was performed \cite{Rainville2003} (blue), are plotted as comparison. The inlet shows our measurement in more detail. The accuracy of the resulting value for $\nicefrac{B_2}{B_0}$ is limited by electric field imperfections.
}
\label{fig:setup}
\end{figure*}

In our trap chamber, we reach a virtually perfect vacuum (better than
$10^{-17}\mbar$), measurable only via the absence of charge exchange of trapped highly charged ions. This is enabled by a pinch-off technique, which hermetically seals the trap-chamber, and cryopumping at $4.2\kelvin$. The sealed trap-chamber makes it necessary to produce the ions inside, which is done using a miniature electron beam ion source (mEBIS) \cite{Alonso2006}. There, electrons are emitted from a field emission point, pass through a hole in our target and are reflected back and forth inside the mEBIS. Due to space charge the electron beam widens and finally hits the target, where atoms and molecules are ablated. The target is made of a plastic compound  with carbon nanotubes to ensure electrical conductivity (TECAPEEK) \cite{Tecapeek}. On top of the target surface, we put a printed layer of deuterated molecules, to allow efficient production of deuteron atomic and molecular ions.
Therefore deuterated thymidine was dissolved in heavy water. From this solution (concentration $5\,\si{\milli\gram\per \milli \litre}$) we put a total of $458$ drops of $5\,\si{\nano\litre}$ each in two layers in an octagonal pattern on top of our target using a drop-on-demand printing system \cite{Haas2017}, see also \autoref{fig:setup}. 

The main systematic limitation in our past experiments has been the residual quadratic component of the magnetic inhomogeneity:
the field generated by the superconducting magnet at the position of our trap, while being very homogeneous, has a slight (in our case positive) quadratic dependency on the axial position. Hence, when an ion moves with its thermal amplitude, it experiences an on average higher magnetic field compared to a situation without the inhomogeneity. This field then depends on the temperature mainly of the axial motion.
We reduced this effect by implementing a closed loop superconducting coil, placed directly around the trap chamber inside our magnet. This coil can be charged from outside, to in-situ shim the magnetic field while simultaneously monitoring the effect on the ion.
With this technique we were able to reduce the quadratic magnetic field inhomogeneity by a factor of $100$ from $\nicefrac{B_2}{B_0}= -7.2(4) \times 10^{-8}\,\si{\per\milli\meter\squared}$ during the proton mass campaign \cite{Heisse2017} to $\nicefrac{B_2}{B_0}= 6.5(6.5)\times 10^{-10}\,\si{\per\milli\meter\squared}$, making the systematic effect on the frequency ratio negligible (see also \autoref{tab:Systematics}).
Construction details on these coils can be found in \cite{Heisse2019}.
To reach a better stability for the frequency ratio $R^{\text{CF}}$, we implemented a pressure stabilization system for the liquid nitrogen (LN2) and liquid helium (LHe) reservoirs. Since the boiling point of cryoliquids is strongly pressure dependent, this potentially reduces magnetic field drifts. With this system we are able to keep the pressure to within $\sim 4\mubar$ of its nominal value, slightly above ambient pressure.
Furthermore, we have added three set screws which can be actuated from the room-temperature stage to allow an exact in-situ alignment of the trap with respect to the magnetic field.
These improvements enabled a stability of $R^{\text{CF}}$ of $1.6\times10^{-10}$ per measured frequency ratio.

\textbf{Data analysis}

For the deuteron mass, the data consists of four different sets of $\text{d}/\isotope{C}^{6+}$ pairs with varied spatial ordering to exclude systematic effects, and in total 41 runs.
The ion pairs were trapped between one to four months and lost due to communication problems between the computer control system and the devices used in the setup.
In each run, settings for the excitation strengths $A_i$ for the PnA remain constant, and one run contains typically 27 frequency ratios.
For each frequency ratio, the ion to be measured first is chosen randomly and shuttled into the precision trap, while the other ion is stored in an adjacent storage trap. After measuring the ion's frequencies, the second ion is transported into the precision trap and the ion which was measured first is parked in a storage trap. The measurement cycle is described in more detail in \cite{Heisse2019}.
The ions are measured using identical trapping potentials, which is possible by tuning the resonator frequency using a varactor diode as described in \cite{Heisse2019}.

To extrapolate energy dependent shifts occurring during the PnA, we took runs with various independently varied excitation strengths for both the deuteron and the carbon ion.
The radius of the modified cyclotron motion $r^+_{\text{exc}}$ is proportional to the excitation strength, $r^+_{\text{i,exc}} = \kappa A_i$, where $\kappa$ is a proportionality constant, which is dependent on the particle.
The energy associated with this motion leads to a relativistic mass increase and therefore to a frequency shift proportional to $\left(r^+_{\text{i,exc}}\right)^2$.
As an example, this shift amounts to $\nicefrac{\Delta\nu_+}{\nu_+}=-19\times10^{-12}$ for carbon and $-18\times10^{-12}$ for deuteron at the lowest excitation amplitude of about $10\mum$ used in the PnA.

\begin{figure}[tb]
\includegraphics[width=0.5\textwidth]{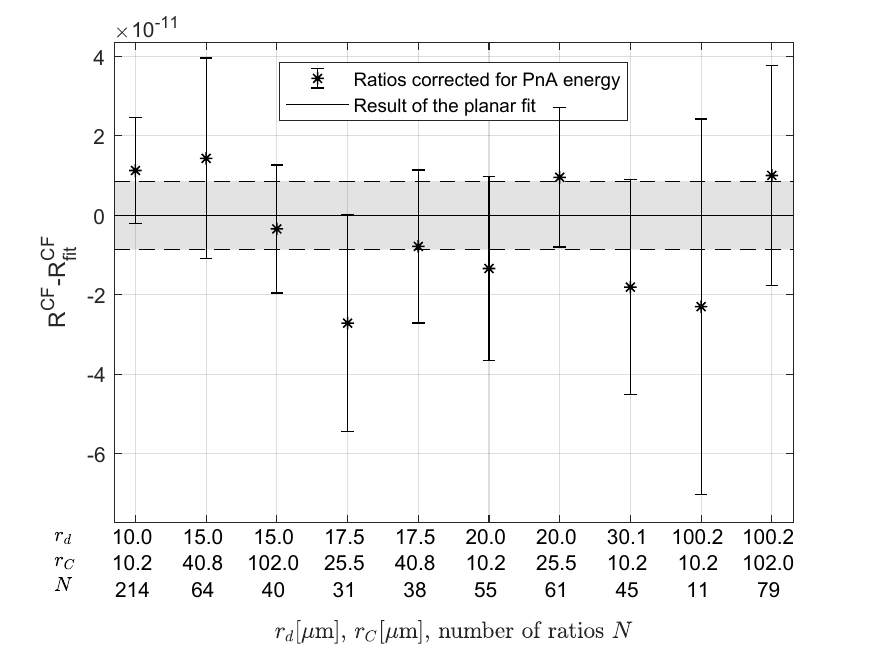}
\caption{\textbf{Averages of cyclotron frequency ratios.} Shown are averages of cyclotron frequency ratios with equal parameters after correction to zero excitation amplitude using the fit described in the text for the first two ion pairs taken with AWG1. Each point corresponds to a setting used in the PnA method. On the x-axis, the corresponding cyclotron radii of deuteron ($r_d$) and carbon ($r_C$) and the number of cyclotron ratios $N$ in each value are given. The error bars denote the standard error of the mean and are estimated from the standard deviations divided by the square root of $N$. The grey band with dashed borders denotes one sigma uncertainty for the fitted frequency ratio.
}
\label{fig:residuals}
\end{figure}

In the analysis this is treated by fitting a plane into the tuples $(A_i(d)^2, A_i(\Csp)^2, R^{\text{CF}})$ in a manner similar to that described in \cite{Heisse2017}. The results of this fit are a ratio extrapolated to zero excitation energy for both ions and denoted as $R^{\text{CF}}_{\text{stat}}$, and by using the known formula for the relativistic shift $\kappa$ as a calibration for the excitation strength.
To further exclude systematic effects, the data with ion pairs three and four have been taken using a different arbitrary waveform generator (AWG) for the excitations. The datasets with the two AWGs have been analysed separately and averaged after.
In \autoref{fig:residuals} the measured ratios for different PnA settings after correction to zero excitation amplitude using the fit are shown for ion pairs one and two. For the other two ion pairs, the data is shown in the supplemental material. Both fit results have been averaged, to arrive at a statistical ratio, which is 
\begin{equation}
R^{\text{CF}}_{\text{stat}}=1.007\,052\,737\,8316\,(54).
\end{equation}

\begin{table}[tb]
\caption{Systematic shifts and their uncertainties after extrapolation to zero excitation amplitude. All values are relative and in parts-per-trillion.
For details see text.}
\label{tab:Systematics}
\begin{center}
\begin{tabularx}{\linewidth}{X | r  r  r  r}
			&	 \multicolumn{2}{c}{Shift of $\nu_c$ for} 	& Correction 	 & Uncer-	\\
	Effect 		&	$d$ & $\Csp$ 				&to $R^{\text{CF}}$			&tainty	 \\ \hline
	Image charge &				$-16.6$&$-98.7$		&	$82.1$ 						&	$4.1$ \\ 
	Special relativity (thermal) & 			$-3.4$&$-0.6$		&	$-2.9$ 						& 	$1.2$ \\ 
	Magnetic inhomogeneity &		$0.4$&$0.1$		&	$0.3$						&	$0.6$ \\ 
	Electrostatic anharmonicity & 	$0$ &		$0$		&	$<0.1$ 						&	$0.3$ \\ 
	Dip lineshape	& 				$0$	&	$0$		&	$0$							&	$4.7$	\\	
	Magnetron frequency	& 		$0$	&	$0$		&	$0$							&	$0.4$	\\ \hline \hline
	Total &					$-19.6$&$-99.2$			&	$79.6$							&	$6.5$
\end{tabularx}
\end{center}
\end{table}

It is important to note that the extrapolation only takes into account the excitation energy.
The modified cyclotron mode was thermalized with the axial tank circuit prior to every PnA cycle to an equivalent axial temperature of $T_{z,FB}=1.2(5)\kelvin$ using electronic feedback cooling. This thermal energy causes a relativistic shift as well, which is treated as a systematic effect.
The largest contribution to the systematic error budget is the image charge shift (ICS). There, the image charge of the ion on the trap surfaces induces an outward force on the ion, effectively decreasing the cyclotron frequency compared to the free space cyclotron frequency. This effect is well understood \cite{Schuh2019}, can be simulated, and corrected for up to an uncertainty caused by the manufacturing precision of the trap. In \autoref{tab:Systematics}, we summarize the systematic corrections to the ratio together with their uncertainties. The systematic effect related to "lineshape" originates from the measurement of the axial frequency using the "dip" technique. There the axial frequency is extracted from a fit to a noise spectrum. The width of this feature is $3\Hz$ for $\Csp$, and the uncertainty given in \autoref{tab:Systematics} corresponds to a frequency uncertainty of $\approx 6\mHz$, which is a line splitting of $\approx 500$. The error bar comes mostly from the determination of the resonance frequency of the tuned circuit used in detecting the axial motion and the frequency pulling associated with it.
The magnetron frequency is also regarded as a source of systematic error because it was not measured every cycle.
Finally, the cyclotron frequency ratio corrected for systematic effects is

\begin{figure}[tb]
\includegraphics[width=0.5\textwidth]{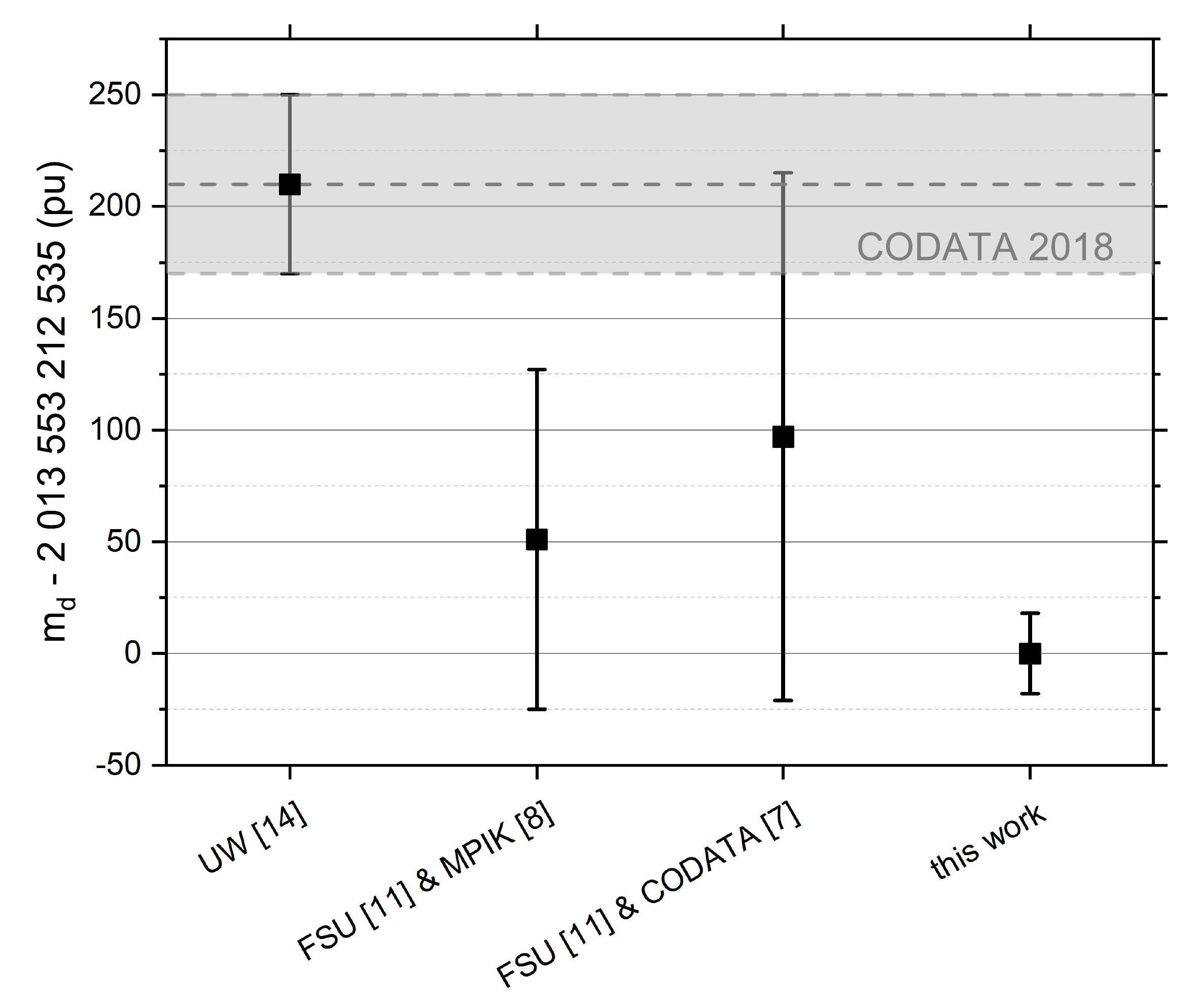}
\caption{\textbf{Most precise mass values for $m_d$.} Shown is a direct measurement by the University of Washington (UW) \cite{Zafonte2015}, the deuteron-to-proton mass ratio by the Florida State University (FSU) \cite{Myers2020}, once combined with the proton mass by this group (MPIK) \cite{Heisse2019}, once combined with the current literature value (CODATA) for the proton mass \cite{CODATA2018}, and the value reported in this work. The CODATA literature value \cite{CODATA2018}, which coincides with the value by the UW since the 2014 adjustment, is shown as a grey band with dashed borders. All error bars correspond to the one sigma confidence interval ($68\%$).}
\label{fig:mDPlot}
\end{figure}

\begin{equation}
R^{\text{CF}}_{\text{final}}=1.007\,052\,737\,9117\,(54)_{\text{stat}}(65)_{\text{sys}}(85)_{\text{tot}},
\end{equation}

where the brackets denote the statistical, systematic and combined uncertainty, respectively. Using \autoref{eq:eq1}, we derive a deuteron mass
\begin{equation}
m_d=2.013\,553\,212\,535\,(11)_{\text{stat}}(13)_{\text{sys}}(17)_{\text{tot}}\amu,
\end{equation}

From this we can also deduce the atomic deuterium mass by adding the electron mass and binding energy: $m(\isotope{D})=2.014\,101\,777\,842\,(17)\amu$.

The same measurement sequence was also used to measure $m\left(\HDp\right)$, which can be used as a stringent test of our systematics.
There we used $\Cfp$ as reference ion, $m(\Cfp)=11.997\,805\,839\,274\,83\,(34)\amu$ and only one pair of ions, which was trapped for seven weeks and removed from the trap on purpose.
We arrive at a statistical ratio
$R^{\text{CF}}_{\text{stat}}=1.007\,310\,263\,850\,(19)$.
The first excited ro-vibrational state of $\HDp$ has a lifetime of $149\secs$ and a transition energy of $5.4\meV$ \cite{Olivares2013}, corresponding to a temperature of $63\kelvin$. Since the mass measurements started several days after the preparation of the $\HDp$ ion, it can be assumed to be in its ro-vibrational groundstate in our $4.2\kelvin$ environment.
Compared to the deuteron measurement, when using a molecular ion one also has to take into account the polarizability \cite{Schiller2014, Thompson2004}. There, the motional electric field induces an electric dipole pointing towards the center of the cyclotron motion. The energy of this dipole in the motional electric field gives rise to an increased effective mass in radial direction, therefore shifting the extracted cyclotron frequency of the $\HDp$ ion in the groundstate by 
$\nicefrac{\Delta \nu_c}{\nu_c} = -1.84 \times 10^{-11}$ compared to a hypothetical particle with the same charge and mass, but no polarizability.
While the electronic spin does not thermalize on the timescale of the experiment, the corresponding shift is less than $0.1$ ppt and therefore negligible.
The final ratio corrected for systematics is
$R^{\text{CF}}\left(\Cfp/\HDp\right)=1.007\,310\,263\,905\,(19)_{\text{stat}}(8)_{\text{sys}}(20)_{\text{tot}}$,
and the mass of the molecular ion
$m\left(\HDp\right)_{\text{direct}}=3.021\,378\,241\,561\,(56)_{\text{stat}}(24)_{\text{sys}}(61)_{\text{tot}}\amu$.

\textbf{Discussion}

The deuteron mass reported here deviates significantly from the current CODATA literature value, see \autoref{fig:mDPlot}. To further validate our measurement, we can compare the directly measured mass of $\HDp$ with the one derived from its constituents.
Using our previously reported proton mass \cite{Heisse2019}, the deuteron mass reported in this work, the electron mass \cite{Sturm2014}, and the binding energy of the three-body system \cite{Korobov2017} one arrives at the following value for the mass of the $\HDp$ molecular ion:
$m(\HDp)_{\text{p+d}}=3.021\,378\,241\,576\,(37)\amu$.
This value agrees with the directly measured one on a one sigma level,
$m(\HDp)_{\text{p+d}}-m(\HDp)_{\text{direct}}=15\,(71)\pamu$.
If one uses the deuteron mass reported by the UW \cite{Zafonte2015}, this difference amounts to $225\,(80)\pamu$, or $517\,(158)\pamu$ if one also uses their previously reported proton mass \cite{VanDyck1999}.
We regard the striking agreement among our measurements with different masses and systematics as a profound consistency check, substantiating our measurement methods.
Using our measurements we can also extract a deuteron-to-proton mass ratio $m_d/m_p (\text{LIONTRAP})=1.999\,007\,501\,228\,(59)$. This value agrees with a direct measurement using $\isotope{H}_2^+$ molecular ions recently reported by the Florida State University \cite{Myers2020} on a one sigma level.

The agreement between the measurements of $m_p$ \cite{Heisse2019}, $m_d$ and $m(\HDp)$ (this work), as well as the deuteron-to-proton mass ratio \cite{Myers2020} opens up the possibility for a least square adjustment of $m_p$ and $m_d$ with only the listed measurements as input and using the techniques as described in \cite{Huang2018}.
The resulting masses are listed in \autoref{tab:masses}. The main advantage of this adjustment is a reduction of the uncertainty of the proton's atomic mass by a factor of $2$ compared to the direct measurement.
With these adjusted masses and the measured deuteron nuclear binding energy $S_n= 0.002\,388\,170\,08\,(42)\amu$ \cite{Dewey2006}, which was adjusted with an updated value for the lattice constant for the crystal used in the measurement (see also supplemental material) \cite{Kessler2017}, the neutron mass becomes
$m_n=1.008\,664\,916\,04\,(42)\amu$, which ist shifted by $9\times 10^{-11}\amu$ compared to the CODATA 2018 literature value \cite{CODATA2018}.
The uncertainty however remains the same due to the limitation by the measurement uncertainty of $S_n$.
When using the adjusted values in this work, the light ion mass puzzle reduces to $\Delta_{\text{FSU}}-\Delta_{\isotope{C}}=258(86)\pamu$, strengthening the credibility of the mass difference between the tritium and  $\text{helium-3}$ nuclei needed for KATRIN.
With the measurements reported in \cite{Heisse2019} and this work, two of the three masses contributing to $\Delta_{\isotope{C}}$ have now been reported from our LIONTRAP collaboration. Together with the remaining $3\,\sigma$ tension in $\Delta$, this is a clear motivation for an independent measurement of the mass of $\text{helium-3}$.

\begin{table}[tb]
\caption{The masses of the light nuclei after adjustment, together with their uncertainties. The resulting value for $\Delta_{\isotope{C}}$ and the value $\Delta_{\text{FSU}}$ are added for convenience.
The values are correlated, with correlation coefficients $r(m_p,m_d)=0.26$, $r(m_p,m_n)=0.03$ and $r(m_d,m_n)=-0.03$
}
\label{tab:masses}
\begin{center}
\begin{tabularx}{\linewidth}{X X c}
	nucleus 	&	value $(\si{\atomicmassunit}$)  & rel. unc. $(10^{-11})$ \\ \hline
	proton mass $m_p$	&	$1.007\,276\,466\,580\,(17)$	&	$1.7$\\
	deuteron mass $m_d$ &	$2.013\,553\,212\,537\,(16)$	&	$0.8$\\ 
	neutron mass $m_n$ 	&	$1.008\,664\,916\,04\,(42)$	&	$42$ \\ \hline
	$\Delta_{\isotope{C}}$ 			&	$0.005\,897\,432\,449\,(50)$&\\
	$\Delta_{\text{FSU}}$ 			&	$0.005\,897\,432\,191\,(70)$&\\
\end{tabularx}
\end{center}
\end{table}

\clearpage

\clearpage
\textbf{Methods}

\textbf{Additional information on the deuteron measurement}

As described in the paper, the data for the deuteron measurement campaign has been taken using two different arbitrary waveform generators (AWGs) for the exciations.
The used generators were AWG1: Keysight 33600A 80 MHz and AWG2: Agilent 33522A 30 MHz.
Since the excitations from two different AWGs in general will not be exactly equal in amplitude, we decided to treat this by taking both datasets separately. In the paper, the data fitting routine is described and the fit for the first two ion pairs is shown.
In extended data figure 1\textbf{a} we show the residuals of the fit used for the second two pairs of ions.

Both fits give ratios extrapolated to zero exciation amplitudes. To combine both fits, which agree very well, a weighted average was used. This is the statistical  ratio $R_{\text{stat}}$ given in the  paper. The single fit results are:
\begin{equation}
\begin{aligned}
\text{AWG} 1 &= 1.007\,052\,737\,831\,3\,(86)\\
\text{AWG} 2 &= 1.007\,052\,737\,831\,7\,(70)\\
\end{aligned}
\end{equation}
The AWG was exchanged because we noticed an increased noise level during double-dip measurements, which might lead to heating. The PnA method is largely unaffected by this, since the excitation is on only for a very short time there, and also much weaker than for a double-dip. We tried to induce a detectable effect by using excessively strong excitations for long times, but did not detect any relevant heating.
The values obtained with the two AWG agree, indicating no systematic influence.

The temperatures of the ions have been measured using standard methods. During the PnA, electronic feedback was used to achieve a temperature $T_{\text{FB}}=1.2\,(5)\kelvin$ for both ions. During the dip measurement, no electronic feedback was used, the temperature was $T_{\text{noFB}}=3.7(5)\kelvin$.

\textbf{Details on the $\HDp$ measurement}

Here we present some more information on the direct measurement of $\HDp$. In principle, the measurement was very similar to the deuteron mass campaign. However, the charge-to-mass ratio of $\HDp$ and $\Cfp$ is $\frac{q}{m}\approx\frac{1}{3}\frac{\si{\elementarycharge}}{\si{\atomicmassunit}}$, resulting in a reduced signal and  a reduced cyclotron frequency of about $\approx 20\MHz$. The trap voltage was adjusted in a manner that approximately the same axial frequencies as in the deuteron measurement were used. The relatively low charge-to-mass ratio gave rise to an extremely narrow dip signal for $\HDp$ of about $0.35\Hz$.

The data of the corresponding surface fit are shown in extended data figure 1\textbf{b} similar to the fits for $m_d$.

In Extended Data Table 1, the systematic shifts and uncertainties for the $\HDp$ measurement are summarized. They differ from the values given for $m_d$ due to the different frequencies and the different charge-to-mass ratio. The temperatures were equal to the temperature in the deuteron measurement.

\textbf{Deuteron-to-proton mass ratio}

In the paper, the deuteron-to-proton mass ratio
\begin{equation}
m_d/m_p (\text{LIONTRAP})=1.999\,007\,501\,228\,(59)
\end{equation}
is given. This ratio is taken from a least square adjustment using the measurments on $m_p$ \cite{Heisse2019}, $m_d$ and $m\left(\HDp\right)$ (this work).
This results in
\begin{align}
m_p&=1.007\,276\,466\,595\,(29),\\
m_d&=2.013\,553\,212\,534\,(17),
\end{align}
with correlation coefficient $r(m_p,m_d)=-0.13$.
The quotient of above values gives the reported deuteron-to-proton mass ratio.

\textbf{Deuteron nuclear binding energy}

The nuclear binding energy of deuteron was measured with Bragg spectroscopy \cite{Kessler99}.
To translate the measured angle into a wavelength, the lattice constant of the silicon crystal used for the measurement is needed.
The lattice constant of the crystal used in the deuteron nuclear binding energy ( ILL2.5 ) has been remeasured in 2006 \cite{Dewey2006} and in 2017 \cite{Kessler2017}.
In this chapter, the 2006 value is updated with the 2017 lattice constant.

The wavelength extracted from the angle measurements in Bragg spectroscopy is proportional to the lattice constant \cite{Dewey2006}. Therefore, to adjust the value for the measured wavelength to a new value of the lattice constant, we use:

\begin{equation}
\lambda_{\text{new}}^{\text{meas}}=\frac{d_{\text{new}}}{d_{\text{old}}}\lambda_{\text{old}}^{\text{meas}}
\end{equation}

Using $d_{\text{new}} = 192.015\,572\,1\,(6\,4) \times 10^{-12}\meter$ \cite{Kessler2017}, $d_{\text{old}}=192.015\,582\,2(9\,6)\meter$ and $\lambda_{\text{old}}^{\text{meas}}=0.557\,671\,328\,(99)\times10^{-12}\meter$ \cite{Dewey2006}, one arrives at
 
\begin{equation}
\lambda_{\text{new}}^{\text{meas}}=0.557\,671\,299\,(97)\times10^{-12}\meter,
\end{equation}

corresponding to an energy of

\begin{equation}
E^{\text{meas}}=2\,223\,248.69\,(39)\eV.
\end{equation}

Note that the error did not change significantly, since the lattice constant measurement was not the dominating contribution to the error.
The measured energy was corrected for nuclear recoil to obtain the nuclear binding energy $S_n$ \cite{Kessler99}:

\begin{align}
S_n&=E^{\text{meas}}+\frac{(E^{\text{meas}})^2}{2 m_d \mathrm{c}^2}\\
	&=2\,224\,566.35\,(39)\eV\\
	&=2.388\,170\,08\,(42)\times 10^{-3}\amu.
\end{align}

Here, $m_d \mathrm{c}^2$ is the atomic mass of deuteron expressed in $\si{\electronvolt}$ using the conversion factor
$(1\si{\atomicmassunit})c^2=9.314\,941\,024\,2\,(2\,8)\times 10^8\eV$ \cite{CODATA2018}.
Note that the discrepancies in $m_d$ as well as the conversion to $\si{\electronvolt}$ do not play any role at the level of precision needed for the recoil correction.

\textbf{Constants}

Where not stated otherwise, the values given by CODATA 2018 \cite{CODATA2018} have been used. In Extended Data Table 2 some of the most important constants used in this work are compiled for convenience.

\textbf{Data availability}

The datasets analysed for this study will be made available on reasonable request.

\textbf{Code  availability}

The analysis codes will be made available on reasonable request.

\textbf{Acknowledgements}
This article comprises parts of the PhD thesis work of S.R.
We acknowledge fruitful discussions on the nuclear binding energy with Michael Jentschel.
This project has received funding from the Max-Planck Society, from the International Max Planck
Research School for Precision Tests of Fundamental Symmetries
(IMPRS-PTFS) and Quantum Dynamics (IMPRS-QD), by the
Max Planck, RIKEN, PTB Center
for Time, Constants and Fundamental Symmetries and
by the Helmholtz Excellence Network ExNet020, Precision Physics, Fundamental Interactions and Structure of Matter (PRISMA+) from the Helmholtz Initiative and Networking Fund.

\textbf{Author contributions} 
The experiment was performed by S.R., F.K., and S.Sa.
The data was analysed by S.R, F.K. and S.St.
The manuscript was written by S.R.
The deuterated target was prepared by R.H., D.R. and C.E.D.
All authors discussed and approved the data as well as the manuscript.

\clearpage
\onecolumn
\includegraphics[width=\textwidth]{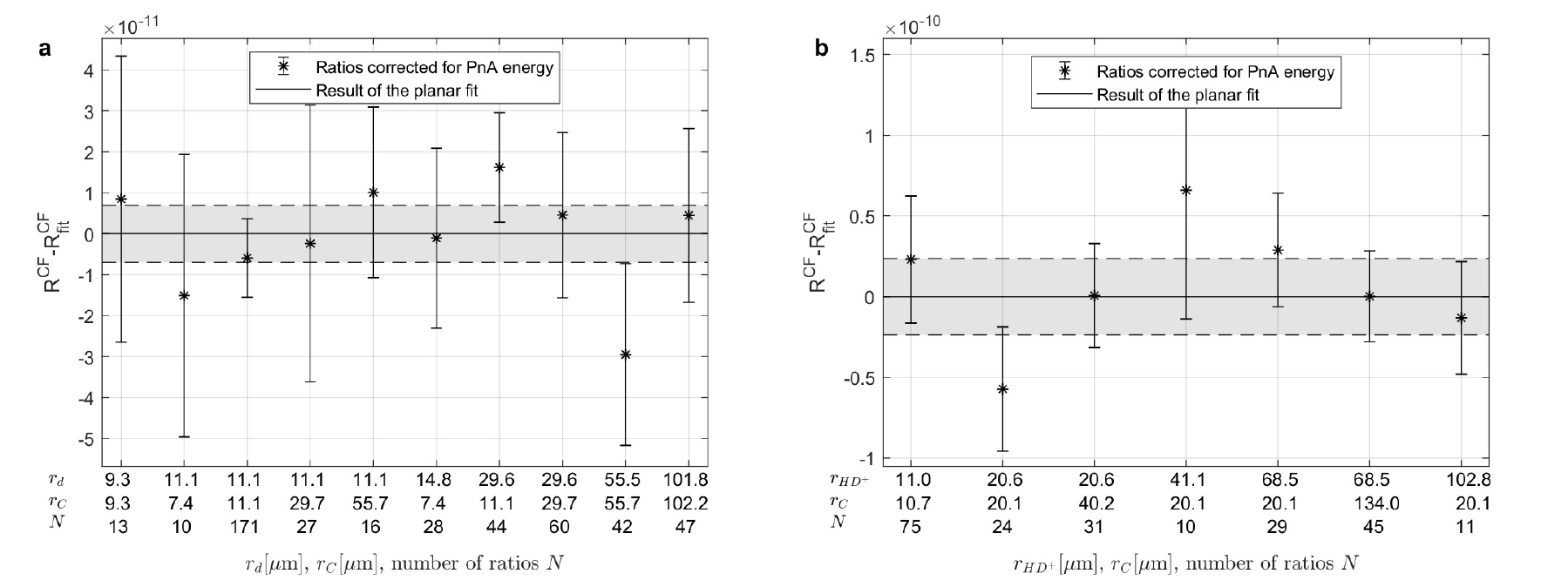}
\textbf{Extended Data Fig. 1: Averages of cyclotron frequency ratios.} Shown are averages of cyclotron frequency ratios with equal parameters after correction to zero excitation amplitude using the fit described in the text for the data with \textbf{a} AWG 2 of the deuteron campaign and \textbf{b} $\HDp$. Each point corresponds to a setting used in the PnA method. On the x-axis, the corresponding cyclotron radii of deuteron ($r_d$) and carbon ($r_C$) and the number of cylotron ratios $N$ in each value are given. The error bars denote the standard error of the mean and are estimated from the standard deviations divided by the square root of $N$. The grey band with dashed borders denotes one sigma uncertainty for the fitted frequency ratio.

\clearpage
\twocolumn
\textbf{Extended Data Table 1: Systematics $\HDp$.} Systematic shifts and their uncertainties after extrapolation to zero excitation amplitude for $\HDp$. All values are relative and in parts-per-trillion.
\includegraphics[width=0.5\textwidth]{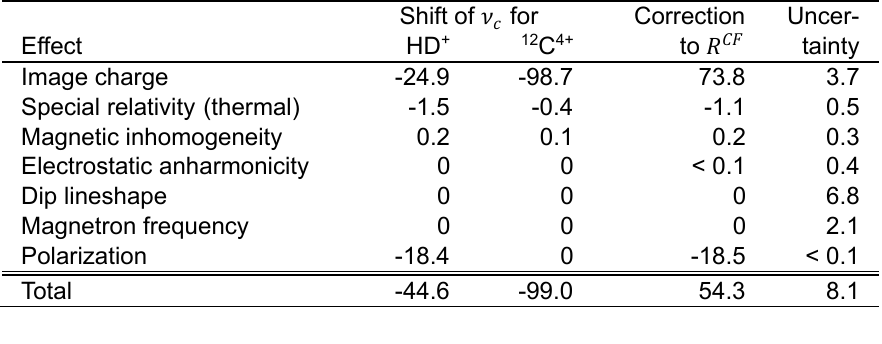}

\clearpage

\textbf{Extended Data Table 2: Constants used in this work.} When no uncertainty is given, the precision of the value is much better than needed for this work. When multiple values are taken from the same reference, the reference is only cited with the first value.
\includegraphics[width=0.5\textwidth]{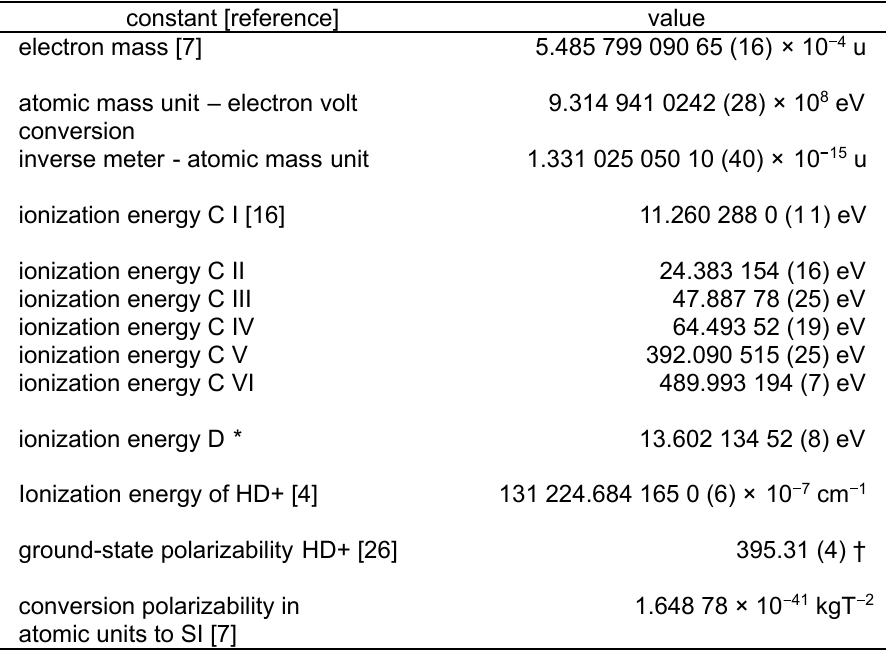}
\end{document}